\documentclass[aps,twocolumn]{revtex4}
\usepackage{bm}
\begin{document}

\title{Comment on ``Silver nanoparticle array structures that produce
  remarkably narrow plasmon lineshapes'' [J. Chem. Phys. 120, 10871
  (2004)]} 
 
\author{Vadim A. Markel}

\affiliation{Department of Radiology, University of Pennsylvania,
  Philadelphia, PA 19104}

\date{\today} 
\maketitle

Recently, Zou, Janel and Schatz (referred to as ZJS below) have
described remarkably narrow plasmon resonances in linear arrays of
silver nanospheres~\cite{zou_04_1}. Without questioning the novelty
and significance of these results, I would like to point out that the
above-referenced paper contains two incorrect statements.

{\bf The first statement} is about my previous work.  Namely, ZJS
write that in a previous study~\cite{markel_93_1} I have considered
``\ldots infinite one-dimensional (1D) arrays in the quasistatic
approximation''. In fact, there was no quasistatic approximation made
in Ref.~\cite{markel_93_1}. The approximation that was made was the
{\em dipole approximation}. These two approximations are distinctly
different. For example, even in the electrostatic limit, the dipole
approximation is grossly inaccurate for two touching conducting
spheres excited by a constant external electric field parallel to the
axis connecting the spheres' centers~\cite{mazets_00_1}. On the other
hand, electromagnetic interaction of small impurities in a crystal or
of dye molecules in large molecular aggregates~\cite{keller_86_1} can
not be understood within the quasistatics, although the dipole
approximation may be very accurate in this case.

Perhaps, the source of confusion is that in Section 2.1 of
Ref.~\cite{markel_93_1} I wrote ``The object under investigation is a
linear infinite chain with step $a$ consisting of point-like dipole
units (monomers)\ldots''. Also, in the introduction of
Ref.~\cite{markel_93_1}, I have suggested that the physical system to
which the considered model is applicable is a molecular aggregate.
Later, in Section 5, I have considered a particular example in which
the polarizability of a dipole, $\alpha$, was given by the quasistatic
polarizability of a small sphere with the appropriate radiative
correction. However, the theoretical formalism of
Ref.~\cite{markel_93_1} did not put any restrictions on $\alpha$. And,
regardless of the form of $\alpha$, the interaction of dipoles was
described with full account of retardation effects.  

In fact, ZJS also work in the dipole approximation, although they
validate their results by comparison with a more general T-matrix
solutions. The situation is somewhat more complicated, however,
because ZJS  use, in addition, an approximation proposed
by Doyle in 1989~\cite{doyle_89_1} in the context of effective-medium
theory of the so-called extended Maxwell-Garnett composites, i.e.,
composites in which inclusions are not small compared to the
wavelength. More specifically, Doyle has studied electromagnetic
properties of a homogeneous host with randomly distributed spherical
inclusions. The essence of the approximation is to consider only
dipole-dipole interactions of the inclusions but to assign them
dynamic dipole polarizability $\alpha$. The latter is given by formula
(\ref{alpha}) below; it is defined as the linear coefficient between
the amplitude of incident plane wave and the total dipole moment of
polarizable sphere of arbitrary size (assuming, the sphere is
isolated) and, in that sense, is exact.  It can be seen that the
Doyle's approach only concerns the choice of $\alpha$ within the
dipole approximation. Thus, it is fully consistent with the general
formalism developed in Ref.~\cite{markel_93_1}.

It should be noted that the accuracy and limits of applicability of
the Doyle's approximation have not been systematically investigated.
In one critical study of extended Maxwell-Garnett
composites~\cite{ruppin_00_1} Ruppin has shown that the Doyle's
approximation is consistent with the asymptotes obtained in the limit
of small volume fraction of inclusions, and, in that limit, allows one
to consider inclusions with size parameters of at least $x\sim 0.5$.
Thus, the Doyle's approximation can be useful for moderate size
parameters. However, if the spherical inclusions are in close
proximity of each other, the secondary scattered waves incident upon
each of them are no longer plane waves. But the dynamic polarizability
used by Doyle is exact only with respect to incident plane waves.
Besides, coupling of higher multipole modes excited in spherical
inclusions can become significant. Therefore, it is quite obvious
that the use of Doyle'a approximation does not fix, in principle, the
deficiencies of the dipole approximation.

{\bf The second statement} concerns the possibility of cancellation of
the imaginary part of denominator in the expression $P=\alpha
E_0/(1-\alpha S)$ (Eq.~5, or, in a more specific form, Eq.~(7) of
Ref.~\cite{zou_04_1}).  This is discussed on p.~10874 of
Ref.~\cite{zou_04_1}. ZJS consider the case when the
incident wave vector is perpendicular to a linear chain of polarizable
dipoles with the period $D$. The polarization of the incident wave is
also perpendicular to the chain. It is stated that the resonance
width, which is related in Ref.~\cite{zou_04_1} to the imaginary part
of the denominator of the above equation, vanishes when $\gamma \leq
8\pi^3A/D^3$, where $\gamma$ and $A$ are parameters which specify the
polarizability of an isolated sphere. Namely, ZJS use the
formula $\alpha = -A/(\omega - \omega_p + i\gamma)$, where $\omega$ is
frequency of incident radiation, $\omega_p$ is the surface plasmon
frequency, and $\gamma$ - the relaxation parameter.  Assuming that the
result ${\rm Im}S=-8\pi^3/D^3$, which is given in Ref.~\cite{zou_04_1}
for $\lambda$ slightly larger than the interparticle distance $D$, is
correct, one immediately can see that the cancellation takes place
exactly at $\gamma=8\pi^3A/D^3$. For smaller values of $\gamma$, the
imaginary part of the denominator becomes, in fact, negative. Such
result clearly contradicts conservation of energy and is unphysical.
It was obtained in Ref.~\cite{zou_04_1} due to several mistakes which
are discussed below.

It is convenient to rewrite Eq.~(5) of Ref.~\cite{zou_04_1} as

\begin{equation}
\label{P_alpha}
P = {{E_0} \over {1/\alpha - S}} \ .
\end{equation}

\noindent
Given the specific form $\alpha=-A/(\omega-\omega_p+i\gamma)$, this
expression differs from Eq.~(7) of Ref.~\cite{zou_04_1} only by
dividing the numerator and denominator by the real constant $A$. The
quantity $S$ here is the ``dipole sum'' - an eigenvalue of the
electromagnetic state of the dipole chain which is excited by incident
radiation. The imaginary part of the denominator of (\ref{P_alpha})
defines total relaxation.

Note that ${\rm Im}(1/\alpha)$ can contain two contributions which
correspond to absorptive and radiative relaxation. Both are strictly
negative. On the other hand, imaginary part of $S$ has nothing to do
with absorptive losses, since $S$ does not depend on material
properties. Thus, ${\rm Im}S$ can only influence radiative relaxation
and can be either positive or negative. In the first case, the
radiative relaxation is increased compared to that of an isolated
sphere, while in the latter case it is reduced. It is important to
note that $1/\alpha$ and $S$ satisfy the following general
inequalities: ${\rm Im}(1/\alpha) \leq
-2k^3/3$~\cite{draine_88_1,markel_92_1} and ${\rm Im}S \geq
-2k^3/3$~\cite{markel_95_1}, where $k=2\pi/\lambda$ is the wavenumber.
Both inequalities follow from the very general consideration of energy
conservation. At the very least, they show that the imaginary part of
the denominator of Eq.~(\ref{P_alpha}) can not become negative. The
radiative relaxation is canceled if ${\rm Im}S=-2k^3/3$ (this
possibility is discussed below).  If, in addition, ${\rm
  Im}(1/\alpha)=-2k^3/3$, total relaxation is equal to zero.
Physically, this can not happen due to small absorption which is
always present even in highly transparent materials, deviations from
the dipole approximation, etc.

Let us re-write the above inequalities for $\lambda\approx D$, which
is the situation considered in Ref.~\cite{zou_04_1}. We obtain ${\rm
  Im}(1/\alpha) \leq -16\pi^3/3D^3$ and ${\rm Im}S \geq
-16\pi^3/3D^3$.  The result adduced in Ref.~\cite{zou_04_1}, namely,
${\rm Im}S=-8\pi^3/D^3\approx -k^3$, clearly contradicts the second
inequality. This is due to two reasons. First, it is incorrect that
the far-field term $\sum_{j\neq i}(k^2e^{ikr_{ij}}/r_{ij})$ dominates
the dipole sum $S$ for $\lambda\approx D$, as is stated in
Ref.~\cite{zou_04_1}. This would be only true for the {\it real part}
of $S$. Second, even if only the far field term is used in the
calculation of $S$, the result adduced in Ref.~\cite{zou_04_1} is off
by the factor of $2$. The correct contribution to ${\rm Im}S$ which
comes from the far-zone term is $[{\rm sgn}(D-\lambda)] 4\pi^3/D^3$.
The contribution which comes from the intermediate-zone term is
$2\pi^3/3D^3$. The contribution from the near-zone term is zero. Thus,
we have ${\rm Im}S=-10\pi^3/3D^3$ for $D<\lambda$ and ${\rm Im}S =
14\pi^3/3 D^3$ for $D > \lambda$ (all calculations are done for
$D-\lambda \ll D$).  It can be seen that the inequality ${\rm Im}S\geq
-16\pi^3/3D^3$ is satisfied {\it strongly}. {\bf Therefore, not only
  the imaginary part of the denominator can not become negative, but
  its exact cancellation is also impossible in the considered
  geometry.} The smallest possible value of ${\rm Im}(1/\alpha - S)$
is equal to $-2\pi^3/3D^3$. However, it is correct that the radiative
relaxation is changed by a significant factor when $\lambda-D$ changes
sign. Thus, $-[16\pi^3/3D^3 + {\rm Im}S] = -10\pi^3/D^3$ for
$\lambda<D$ and $-[16\pi^3/3D^3 + {\rm Im}S] = -2\pi^3/D^3$ for
$\lambda>D$, a drop by the factor of $5$.  This can be practically
important if radiative losses are dominant over absorptive losses.

Next, we discuss the inequality ${\rm Im}(1/\alpha) \leq -2k^3/3$.
This inequality ensures that the dipole contribution to the absorption
cross section of a particle is not negative. It must hold even for
nonabsorbing particles and, in particular, for $\gamma=0$. In the case
of a small particle, this inequality is satisfied if one uses the
quasistatic polarizability with the inclusion of the radiative
reaction correction: $\alpha = \alpha^{(QS)}/(1 -
2ik^3\alpha^{(QS)}/3)$. Here $\alpha$ is the polarizability with the
radiative correction and $\alpha^{(QS)}=R^3(\epsilon-1)/(\epsilon+2)$
is the quasistatic polarizability, $R$ being the sphere radius. The
importance of the radiative correction is discussed, for example, in
Ref.~\cite{draine_88_1}, and the authors of Ref.~\cite{zou_04_1} are
also aware of it (see Ref.~\cite{kelly_03_1}, Eqs.~16-18). The
expression $\alpha = -A/(\omega -\omega_p +i\gamma)$ used in
Ref.~\cite{zou_04_1} does not contain the radiative correction.
Therefore, its use (together with an incorrect expression for $S$)
leads to unphysical results in the limit $\gamma \rightarrow 0$, such
as the total cancellation of relaxation or negative relaxation. It
should be also noted that the dynamic expression for $\alpha$ which
ZJS used in numerical simulations (according to the
Doyle's approximation) also satisfies the above inequality. Indeed, if
we take

\begin{equation}
\label{alpha}
\alpha = {{3i} \over {2k^3}} {{m\psi_1(mkR)\psi_1^{\prime}(kR) -
    \psi_1(kR)\psi_1^{\prime}(mkR)} \over {m\psi_1(mkR)
    \xi_1^{\prime}(kR) - \xi_1(kR)\psi_1^{\prime}(mkR)}} \ ,
\end{equation}

\noindent
where $\psi_1$ and $\xi_1$ are the Riccati-Bessel functions,
$m=\sqrt{\epsilon}$ is the complex refractive index of the spheres,
then the Taylor expansion of ${\rm Im}(1/\alpha)$ in powers of the
wave number reads

\begin{eqnarray}
{\rm Im}(1/\alpha) = -\frac{2 {k^3}}{3} - {{3{\rm Im}\epsilon} \over
  {R^3 \vert \epsilon -1 \vert^2}} - {{3k^2{\rm Im}\epsilon} \over
  {5R\vert \epsilon -1\vert^2}} \nonumber \\ - {{3k^4 R(8 + \vert\epsilon\vert^2 -2
    {\rm Re}\epsilon){\rm Im}\epsilon} \over {350\vert\epsilon -
    1\vert^2}} + O(k^6 R^3) \ .
\label{alpha_exp}
\end{eqnarray}

\noindent
The expansion beyond the third order contains only even powers of $k$
and it can be verified that each term in the expansion is
non-positive. The exact equality ${\rm Im}(1/\alpha)=-2k^3/3$ takes
place only for non-absorbing materials with ${\rm Im}\epsilon=0$
(which do not occur in nature).

Finally, we discuss the possibility of exact cancellation of the
radiative relaxation. Note that (i) only the {\it radiative part} of
relaxation can be zero, (ii) the {\it total relaxation} is always
nonzero due to nonzero absorption, but can become, in principle,
arbitrarily small, and (iii) such cancellation can not take place in
the geometry considered in Ref.~\cite{zou_04_1}. Generally, there can
be two reasons for cancellation of the radiative relaxation.  The
first is symmetry~\cite{markel_95_1}. Within the dipole approximation,
the cancellation takes place when the symmetry of a particular
excitation mode is such that dipole radiation is forbidden.  A
non-zero radiative relaxation can still result from higher-multipole
radiation, similarly to non-zero decay rates of excited atomic states
whose decay is dipole-forbidden. The second reason is when photon
emission is prohibited by conservation laws, such as the light cone
condition~\cite{burin_04_1}. In a linear chain of dipole-polarizable
particles the cancellation of radiative relaxation can take place when
the incident wave vector is parallel to the chain. However, the
radiative relaxation is always nonzero for normal
incidence.

\bibliography{abbrevplain,article} \end{document}